\documentclass[11pt,twoside]{article}
\usepackage{asp2010}

\resetcounters

\begin{document}

\title{Statistical Tools of Interstellar Turbulence: Bridging the Gap Between Numerics and Observations}
\author{B. Burkhart $^1$, A. Lazarian$^1$}
\affil{$^1$ {Astronomy Department, University of Wisconsin, Madison, 475 N.  Charter St., WI 53711, USA}}

\begin{abstract}
MHD Turbulence is a critical component of the current paradigms of star formation, particle transport, magnetic reconnection  and evolution of the ISM.
Progress on this difficult subject is made via numerical simulations and observational studies.  However, due to limitations of resolution,
scale disprencies, and complexity of the observations,  the best approach for connecting numerics to observations is not always obvious.  
Here we advocate for a approach that invokes statistical techniques to understand the underlying physics of turbulent astrophysical systems.
The wealth of numerical and observational data calls for new statistical tools to be developed in order to study
turbulence in the interstellar medium.  We briefly review some of the recently developed statistics that focus on characterizing
gas compressibility and magnetization and their uses to interstellar studies. 
\end{abstract}

\section{Introduction}
	The paradigm of the interstellar medium has undergone major shifts in the past two decades thanks to the combined efforts of high resolution surveys and the exponential
increase in computation power allowing for more realistic numerical simulations.  The ISM is now known to be
highly turbulent and magnetized, which affects ISM structure, formation, and evolution.  

 	To this date, turbulence has always been understood in a statistical manner - showing the 'order from chaos.'
In the ISM in particularly, statistical studies have proven to be important in characterizing  
the properties of the magnetized turbulent ISM  (Lazarian 2009). Indeed, a statistical description
is necessary in any situation where the 
subject of turbulence arises, as it allows one to study the underlying regularities of the fluid motion (Frisch 1995).

	The most common ``go-to'' statistical tool for both
observers and theorist alike is the spatial power spectrum. In fact, most of
the attempts to relate observations to models has been by obtaining the spectral
index (i.e. the log-log slope of the power spectrum) of column density and
velocity. Although the power spectrum is useful for obtaining information about energy
transfer over scales, it does not provide a full picture of turbulence.
This is partially because it only contains information on Fourier amplitudes and neglects information in the phases.
 In light of this, many other techniques have been developed
to study and parametrize observational magnetic turbulence.
 For astrophysical settings in particular, these statistical studies have found their 
uses in the comparison of observations with models
of turbulence. These statistics include probability density functions (PDFs), wavelets, the principal
component analysis,  higher order moments, Tsallis statistics, spectrum and bispectrum 
(Brunt \& Heyer 2002; Kowal, Lazarian \& Beresnyak 2007,  Burkhart et al. 2009, Esquivel \& Lazarian 2010, Toffelmire, Burkhart, \& Lazarian 2011). 
 Wavelets methods, such
as the $\Delta$-variance method, have also been shown to be very useful in characterizing
structure in data (see Ossenkopf et al. 2008a).  

\begin{figure}[tbh]
\centering
\includegraphics[scale=.3]{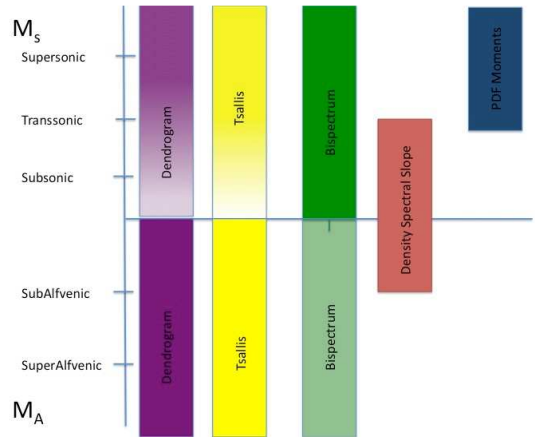}
\caption{Cartoon showing different statistics studied with their dependencies on the sonic and Alfv\'enic Mach numbers.  The different colors indicate
different statistics.  The intensity of the colors indicate the confidence the statistics can provide for the parameter on the y-axis in our simulations.}
\label{fig:ps}
\end{figure}

In general, the best strategy for studying a difficult subject like interstellar turbulence is to use a synergetic approach, 
 combining theoretical knowledge, numerical simulations,
and observational data via statistical studies. 
 In this way one can obtain the most complete and reliable picture of the physics of turbulence.  
Here we seek to extend the statistical comparison between numerical and observational turbulence by reviewing statistical tools that can
greatly compliment the information provided by the power spectrum.   In particular we focus the review on 
tools that can provide information on sonic and Alfv\'enic Mach numbers.  We understand the sonic Mach number
to be defined as  ${\cal M}_s \equiv {\bf v}/C_s$, and the Alfv\'enic Mach number ${\cal M}_A\equiv {\bf v}/v_A $, where 
$v_A = |{\bf B}|/\sqrt{\rho}$ is the Alfv\'enic velocity, ${\bf B}$ is magnetic field and $\rho$ is density. 
 These parameters are not always easy to characterize observationally, with the
Alfv\'enic Mach number being particularly difficult due to cumbersome observational measurements of vector magnetic field.  

This review will highlight several different tools (see Figure 1) studied in the works of Kowal et al. 2007, 
Burkhart et al. 2009, 2010, 2011, Esquivel et al. 2010, and Toffelmire et al. 2011  which represent a mixture of numerical and observational
studies.   We focus on statistics that have application for observable data, in this case, particularly PPV or column density.

\begin{figure*}[tbh]
  \begin{center}
    \includegraphics[keepaspectratio=true,scale=0.3]{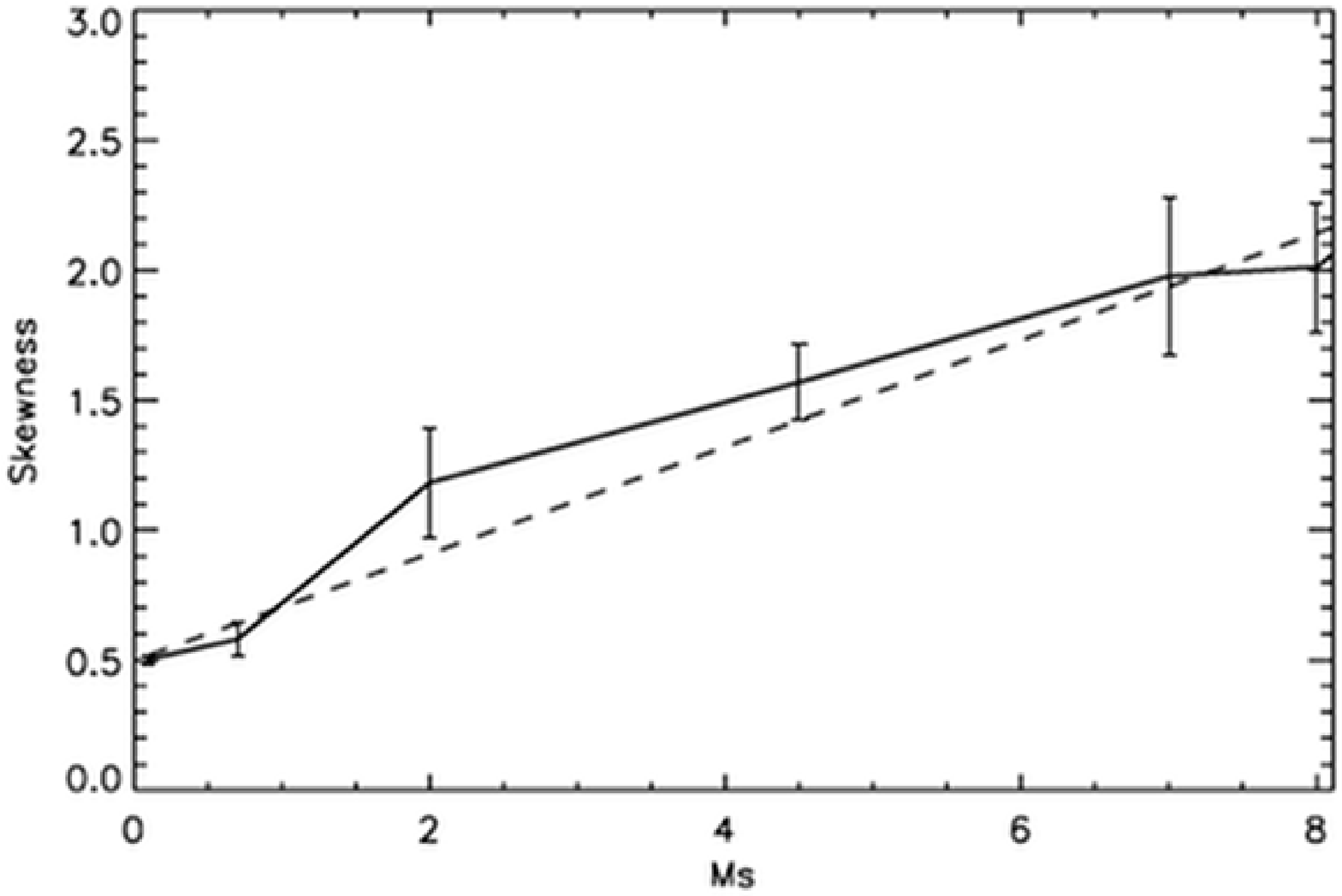}
     \includegraphics[keepaspectratio=true,scale=0.88]{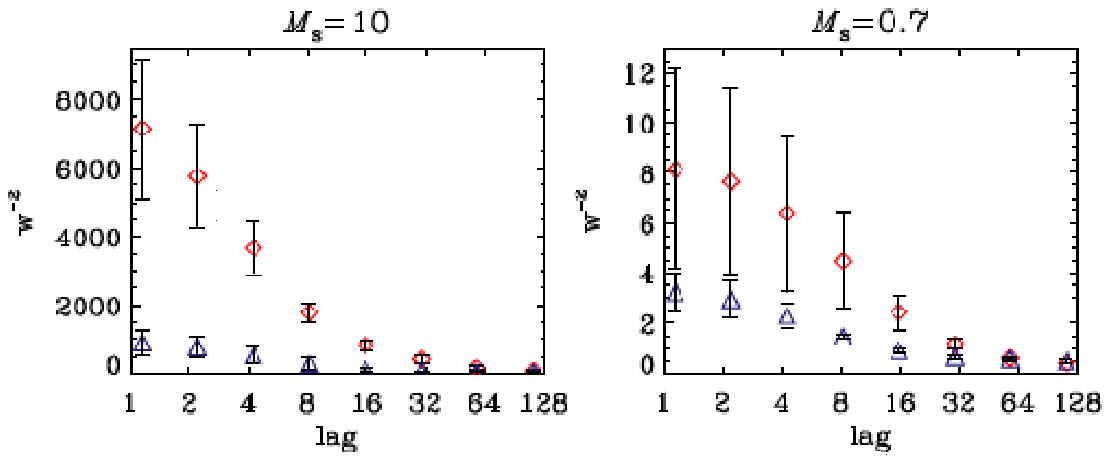}
    \caption{ Uses of PDFs for characterizing turbulence. Top row: Skewness vs. sonic Mach number for six different simulations of MHD turbulence. Similar trends are seen in kurtosis. Dashed line
    is a linear best fit.  From Burkhart et al. 2010
    Bottom row: Tsallis of simulations with  ${\cal
        M}_s$= 10, 7, 4, 0.7 (top left to bottom right respectively).  In this case the field G(x) as shown in 
       equation 2 is 3D density.  We plot the Tsallis fit parameter $w$ vs. lag. 
      Sub-Alfv\'enic simulations are denoted with red diamonds while super-Alfv\'enic 
      simulations are denoted with blue triangles.  From Toffelmire et al. 2011.}  
    \label{fig:3d4}
  \end{center}
\end{figure*}

\begin{figure*}[htb]
\centering
\includegraphics[scale=.4]{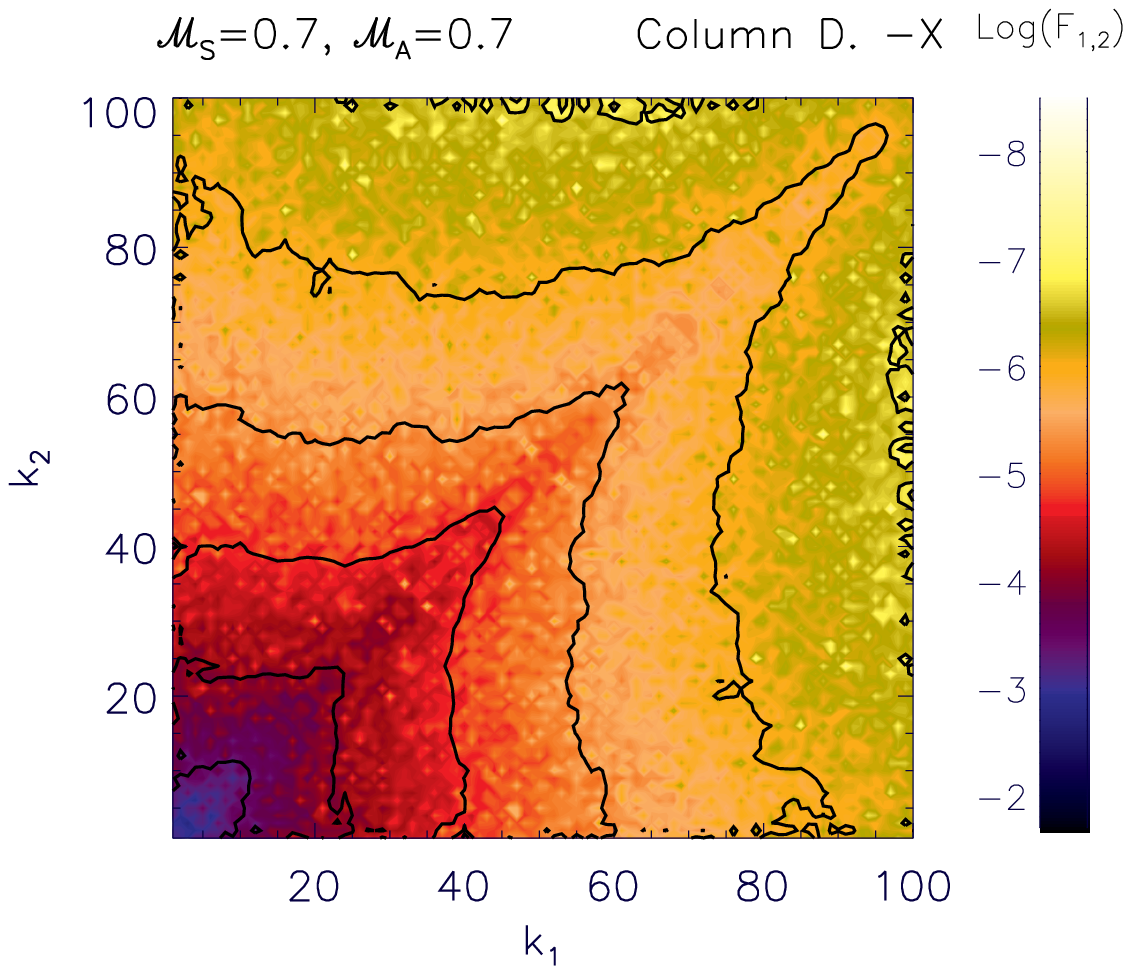} 
\includegraphics[scale=.4]{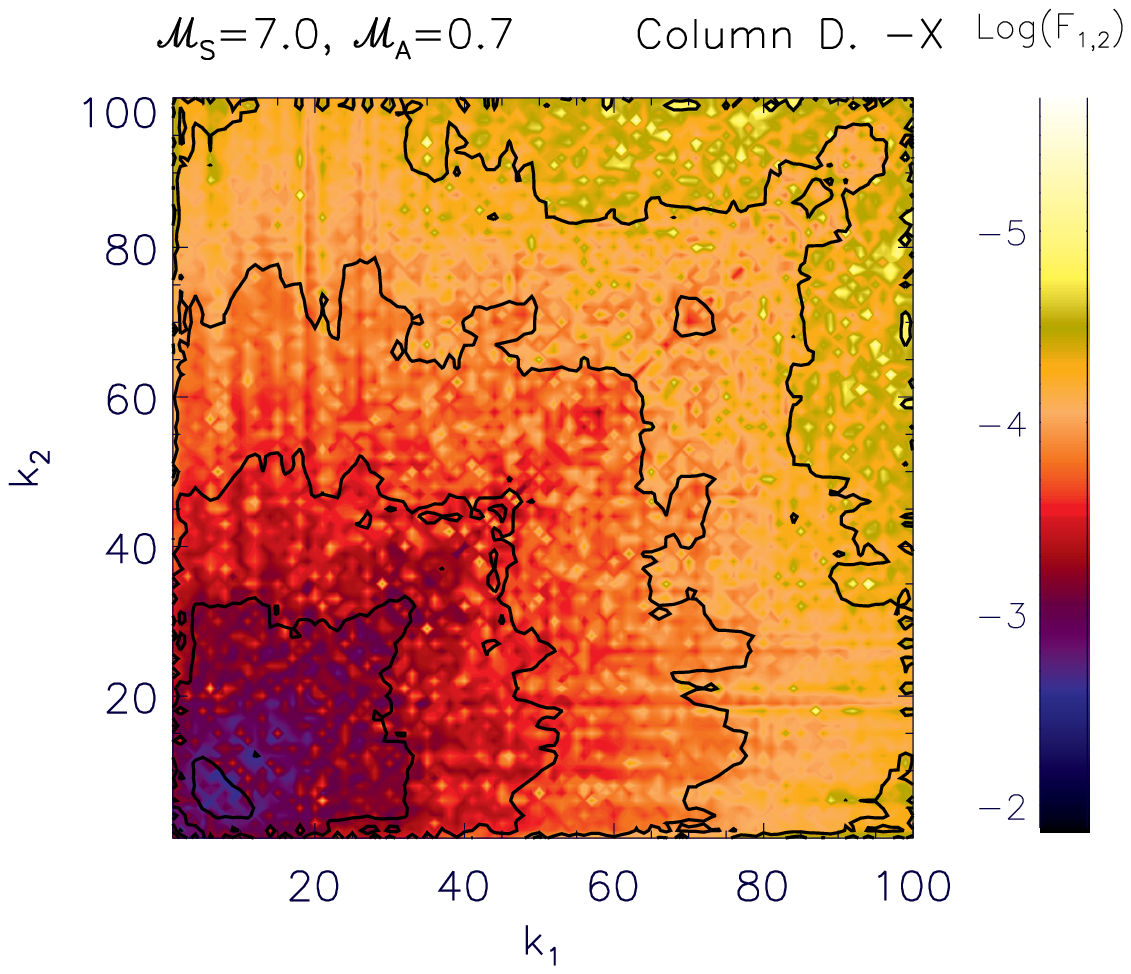} 
\caption{The amplitude of the bispectrum for the scaled simulated
column density. 
The left plot shows a subsonic model, while the right plot is for a supersonic model.  
Both models have ${\cal M}_A$=0.7.  These figures show the degree of correlation 
between wavenumbers $k_{1}$ and $k_{2}$. 
The supersonic model has higher bispectral amplitudes, and more
circular isocontours, therefore a stronger correlation between wave modes. From Burkhart et al. 2010
\label{fig:bispsimms}}
\end{figure*}

\section{The Sonic Mach Number}
The sonic Mach number describes the ratio of the flow velocity to the sound speed, and thus is a measure of the compressibility of the medium. 
Turbulence that is supersonic displays very different characteristics from subsonic turbulence in terms of the spectral slope and density/velocity 
fluctuations.   Because the physical environment of compressible turbulence is very different from incompressible, this parameter is extremely
important for many different fields of astrophysics including, but not limited to, star formation and cosmic ray acceleration.  

\subsection{Higher Order Moments of Column Density}
Moments of the density distribution can be used to roughly determine the gas compressibility through shock density enhancements.  As the ISM
media transitions from subsonic to supersonic and becomes increasingly supersonic, shocks create enhanced intensity, which is reflected in the 
mean value and variance of the
density and column density PDFs.  In addition, as the shocks become stronger, the PDF is skewed and becomes more kurtotic than Gaussian. 
Thus, one can apply very simple statistical descriptors such as skewness  (see Figure 2, top) and kurtosis to astrophysical maps and gain insight into
the compressibility of the region.   However, Kowal et al. 2007 showed
that this method is not very effective for subsonic cases, as these distributions are roughly Gaussian.  For areas where Mach numbers approach and 
exceed unity, higher order moments of column density PDFs can be used as a measure of compressibility.

\subsection{Bispectrum}
While the power spectrum has been used extensively in ISM studies, higher order spectrum have been more rare.  The bispectrum, or Fourier transform of the
3rd order autocorrelation function, has been applied to isothermal ISM turbulence simulations and the SMC only recently (Burkhart et al. 2009, 2010) although
it is extensively used in other fields including cosmology and biology.

The bispectrum preserves both the amplitude and phase and  provides information on the interaction of wave modes.  Completely randomized modes will
show a bispectrum of zero, while mode coupling will show non-zero bispectrum.   Shocks and high magnetic field have been shown to increase mode coupling in the 
bispectrum (Burkhart et al. 2009, 2010). The bispectrum shows a particular sensitivity to picking out shocks in the medium (see Figure 3).
 Due to their ability to shallow out the density energy spectrum, shocks greatly enhance the small scale wave-wave coupling.

\section{Magnetization of Turbulence}
The Alfv\'en number is the dimensionless ratio of the flow velocity to the Alfv\'en speed.   As the Alfv\'en speed depends on the magnetic field, this
ratio can provide information on the strength of the magnetic field relative to the velocity and density.  The Alfv\'enic number is critical in several fields including
interplanetary studies and star formation.  The solar wind is known to be a  super-Alfv\'enic flow while the Alfv\'enic number in star forming
regions is still hotly debated. 
\subsection{Tsallis PDFs of PPV and Column Density}
PDFs of increments (of density, magnetic field, velocity etc.) are a classic way to study turbulence since the phenomena is scale dependent. 
The Tsallis function was formulated in Tsallis 1988  as a means to extend traditional Boltzmann-Gibbs mechanics 
to fractal and multifractal systems.

\begin{figure*}[tbh]
\centering
\includegraphics[scale=.45]{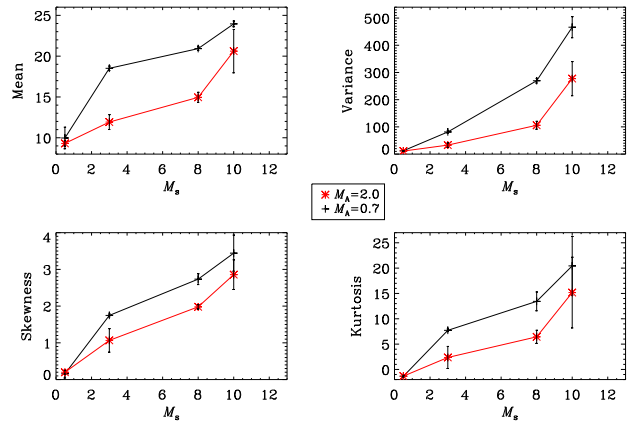}
\caption{Moments of the dendrogram tree (leaves + branches) vs. ${\cal M}_s$ for eight different simulations spanning a range of
sonic numbers from 0.5 to 10.  Here we have chosen $\delta$=4. Each sonic number is divided into sub-Alfv\'enic or super-Alfv\'enic.
 Panels show  mean, variance, skewness and kurtosis of the distribution. Colors indicate different Alfv\'en number. From Burkhart et al. 2011.}
\label{fig:p1}
\end{figure*}

\begin{equation}
R_{q}= a \left[1+(q-1) \frac{\Delta f(x,r)^2}{w^2} \right]^{-1/(q-1)}
\label{eq:1}
\end{equation}

 The Tsallis distribution  (Equation 1) can be fit to PDFs of increments, that is, 
\begin{equation}
 f(x,r)=G(x + r) - G(x)
\end{equation}
\label{eq:e3}
 where G(x) is a particular field (for example, turbulent density, velocity
 or magnetic field) and r is the lag.
The Tsallis fit parameters (q, a, and w in Equation 1) describe the width, amplitude, and tails of the PDF.  These parameters have been shown
to have dependencies on \emph{both} sonic and Alfv\'enic Mach number, however we include Tsallis in this section because it is highly sensitive to the Alfv\'enic Mach number of turbulence.
 Tsallis parameters are able to distinguish between subsonic , transsonic and supersonic
turbulence as well as gauge whether the turbulence is sub-Alfv\'enic or super-Alfv\'enic.  We provide an example  of the fit parameter w (see Equation 1)  in Figure \ref{fig:3d4} bottom row,
taken from Tofflemire, Burkhart, \& Lazarian 2011.   This parameter describes the width of the PDF and is  particularly sensitive to both the Alfv\'enic and sonic Mach numbers. 
The amplitude parameter (not shown) a is sensitive to the sonic Mach number, while the kurtotic q parameter (not shown) is not sensitive to either of the Mach numbers.

\subsection{ Dendrograms of Position-Position-Velocity (PPV) data}
The ISM is fractal in nature (Stutzki 1998), and as such, the observed gas structures
are often hierarchical.  This is especially true in the case of molecular gas or where the sonic number is high and dense filaments
develop. A Dendrogram  is a hierarchical tree diagram that
has been used extensively in other fields, particular
galaxy evolution and biology. It is a graphical representation
of a branching diagram, and for our particular
purposes with PPV data, quantifies how and where local
maxima of emission merge with each other. The dendrogram
was first used on ISM data in Rosolowsky et al.
2008 and Goodman et al. 2009 in order to characterize
self-gravitating structures in star forming molecular
clouds.  The dendrogram picks out local maxima in the data based on a user set threshold value $\delta$, then contours the data and travels
down the contours till it finds a level they merge at.   

Burkhart et al. 2011 used the dendrogram on synthetic PPV cubes (all normalized to unity, which makes the threshold parameter $\delta$ easier to interpret)
 and found it to be rather sensitive to magnetic density/velocity enhancements. In particular, 
they studied how the moments of the tree diagram distribution vary with the sonic and Alfv\'enic Mach number and
level of self-gravity.
These moments showed clear signs
of being dependent on Mach numbers (see Figure \ref{fig:p1} for an example with $\delta=4$).
In addition to being sensitive to the Mach numbers of turbulence
the dendrogram is also able to distinguish between simulations that
show varying degrees of gravitational strength.  

\section{Conclusions}
The last decade has seen major increases in the knowledge of the ISM and of its turbulent nature thanks to high resolution observations and advanced numerical simulations.  
This calls for new advances in statistical tools in order to best utilize the wealth of observational data in light of numerical and theoretical predictions. 
Recently several authors have explored new tools for studying turbulence beyond the power spectrum.   While these proceedings do not cover nearly all the useful
tools in the literature, we attempt to provide some review on tools that describe the gas compressibility and the Alfv\'enic Mach number by utilizing density fluctuations
created by shocks and magnetic density enhancements.

\end{document}